\documentclass{aastex}          
\usepackage{spr-astr-addons}    

\begin{document}
%
\title{The UV window on counter rotating ETGs: insight from SPH  simulations
with chemo-photometric implementation}

\shorttitle{UV view of multi spin galaxies}
\shortauthors{Bettoni et al.}

\author{D. Bettoni\altaffilmark{1}} \author{P. Mazzei\altaffilmark{1}} \and \author{R. Rampazzo\altaffilmark{1}}
\affil{(1) INAF - Osservatorio Astronomico di Padova}
\email{daniela.bettoni@oapd.inaf.it} 
\and
\author{A. Marino\altaffilmark{2}} \and \author{G. Galletta\altaffilmark{2}}
\affil{(2) Dipartimento di Fisica e Astronomia, Universit\'a di Padova}
\email{giuseppe.galletta@unipd.it}
\and
\author{L. M. Buson\altaffilmark{1}} 
\affil{(1) INAF - Osservatorio Astronomico di Padova}
\email{lucio.buson@oapd.inaf.it} 

\begin{abstract}
 The Galaxy Evolution Explorer ({\it GALEX}) detected
 ultraviolet  emission in about 50\% of multi-spin early-type galaxies (ETGs), 
 suggesting  the occurrence of a recent rejuvenation 
episode connected to the formation of these kinematical features.  
With the aim at investigating the complex evolutionary scenario leading to
the formation of counter rotating ETGs (CR-ETGs)   we use our Smooth 
Particle Hydrodynamic (SPH) code with chemo-photometric  implementation. 
We discuss here the UV evolutionary path of two CR-ETGs, NGC 3593  
and NGC 5173,  concurrently best fitting their  global observed properties, 
i.e., morphology, dynamics, as well as their total $B$-band absolute magnitude and spectral energy  distribution 
(SED) extended over three orders of magnitude in wavelength.
These simulations correspond to our predictions about the
target evolution which we follow in the
 color-magnitude diagram (CMD), near-UV ($NUV$) versus $r$-band 
 absolute magnitude, as a powerful diagnostic tool to emphasize 
 rejuvenation episodes.
\end{abstract}


\section{Introduction}
\label{sec:1}
The phenomenon of counter rotation, i.e.  two galaxy  components (gas and/or stars) 
 rotating in opposite directions (see \citet{Gal} and \citet{CB} for a review) is 
 considered one of the most impressive scars of interaction/accretion/merger
 episodes.
Counter rotation has been revealed in galaxies of all morphological types, from ellipticals to spirals, making the phenomenon of general interest. 

The detailed study of the CR components may shed light on 
 the external mechanisms of galaxy formation/evolution, on the complex galaxy 
 response to accretion/merging episodes, as well as on possible inner evolutionary mechanisms.  
Indeed, in some  cases of stars vs. stars counter-rotation in disk galaxies,  
 an internal origin, {\it secular evolution}, is considered viable as a self-induced 
 phenomenon by non-axisymmetric potentials \citep{WP}.  
The current challenge is to disentangle, among  the  manifold of cases 
observed, the different evolutionary paths producing CR
galaxy components and to study the ``side effects'' on the host with a
multi-wavelength approach.

In this scenario, the  UV window has a special value since it is sensitive to the
presence of a young stellar component in  otherwise ``red \& dead'' ETGs
\citep{Rampazzo11}. 
In the past years it became evident that the UV spectra of quiescent ETGs 
can be modeled as a sum of two components: a normal, cool stellar population of main
 sequence and giant stars plus a very blue population having a steeply rising UV flux 
 below 2000 \AA ~\citep{Bur}. 
 More recently, {\it GALEX} showed that a surprisingly high 
 fraction (15\%)   of  ETGs red in the Sloan Sky Survey (SDSS \citet{A09}) exhibit strong UV 
 excess \citep{Yi}. Later, \citet{D07} and \citet{SK07} showed that up to ~30\% of the 
 ETGs imaged with  {\it GALEX} show signatures of rejuvenation 
 episodes, even after having excluded  classical UV-upturn candidates.
\citet{R07} and \citet{M09} showed that ETGs with shell structures 
(indicative, according  to simulations, of recent accretion episodes) host a
 {\it rejuvenated} nucleus. Similar results have been obtained by \citet{J09} 
 for the {\tt SAURON} galaxy sample \citep{Zee02}.  \citet{J09}  and \citet{M09} detected also UV bright outer rings.
 \citet{Marino11} showed that such rings consist of young ($<$200 Myr old) stellar populations, accounting for
 up to 70\% of the FUV flux but containing only  few percents of the total stellar mass. 
In addition, the distribution of galaxies in the (NUV-r) vs $r$-band absolute magnitude diagram 
\citep[e.g.][]{Baldry04, Lewis02, Strateva01} evidences that the red galaxy population is the result of
transformations, mainly driven by the  environment, of the blue, late-type galaxies 
via mechanisms of star formation quenching. The galaxy transition from star forming to quiescent
galaxies, is captured by the presence on CMDs of an intermediate zone,
the so--called {\it Green Valley} (GV) \citep{Martin07, Wyderetal07}.  
Investigating galaxies in the GV should shed light on the mechanisms governing 
the {\it on-off} state of the star formation \citep{MP14a, MP14b}.

Here we match the global observed properties and predict the evolution
 of  two CR-ETGs, NGC 3593 and NGC 5173, using
 SPH simulations with chemo-photometric implementation.\\
  The plan of the paper is as follows.
In section \ref{sec:2} we summarize the recipes of our SPH models  widely described in previous papers
\citep[][and references therein]{MP14a,MP14b}; in section \ref{sec:3} we describe the 
sample of CR galaxies we plan to study, and the results for two
cases of interest, i.e., NGC 3593 and NGC 5173; in section \ref{sec:4} we drawn our  conclusions. 

\begin{figure*}[!]
\includegraphics[scale=.39]{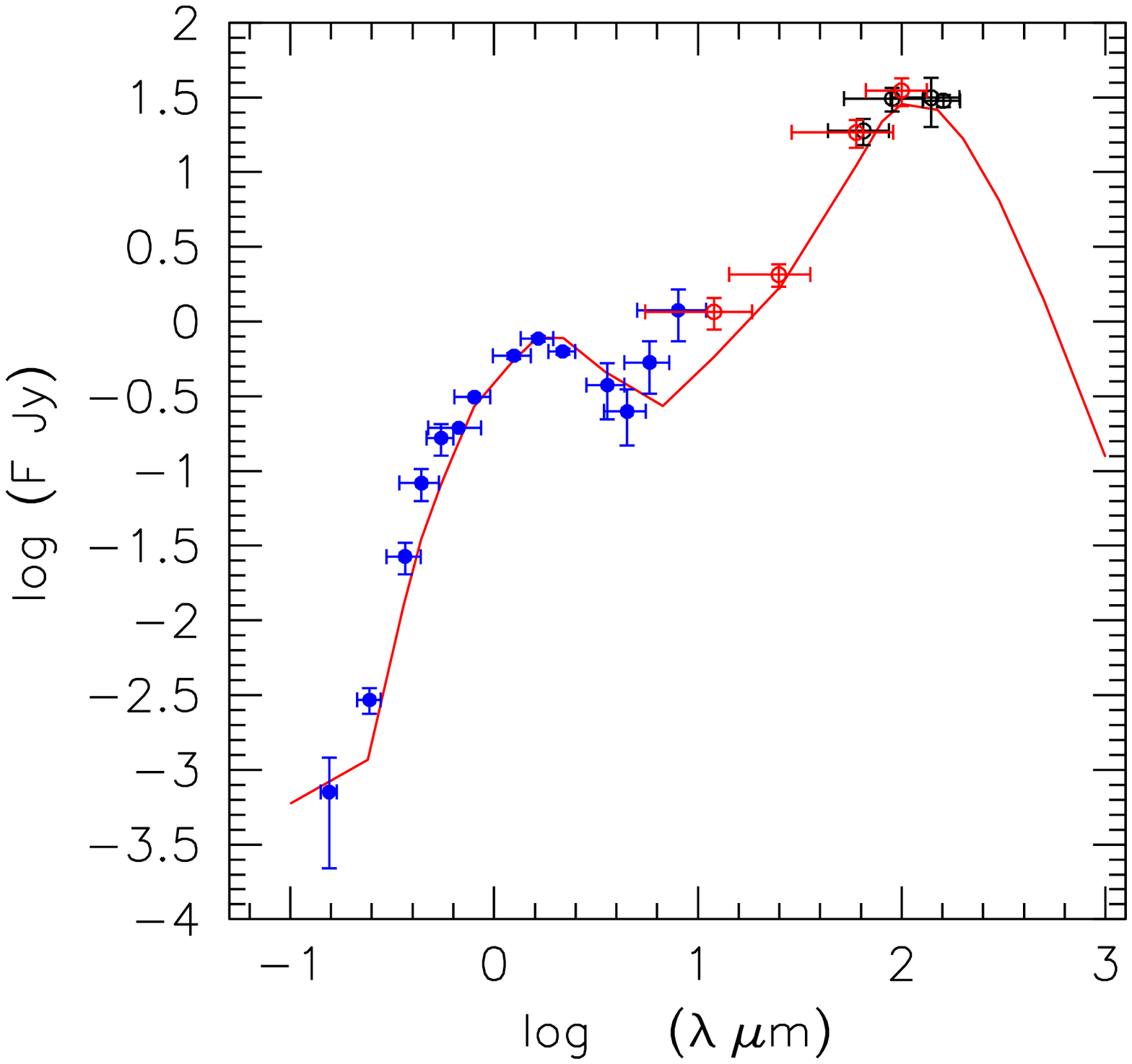}
\includegraphics[scale=.32]{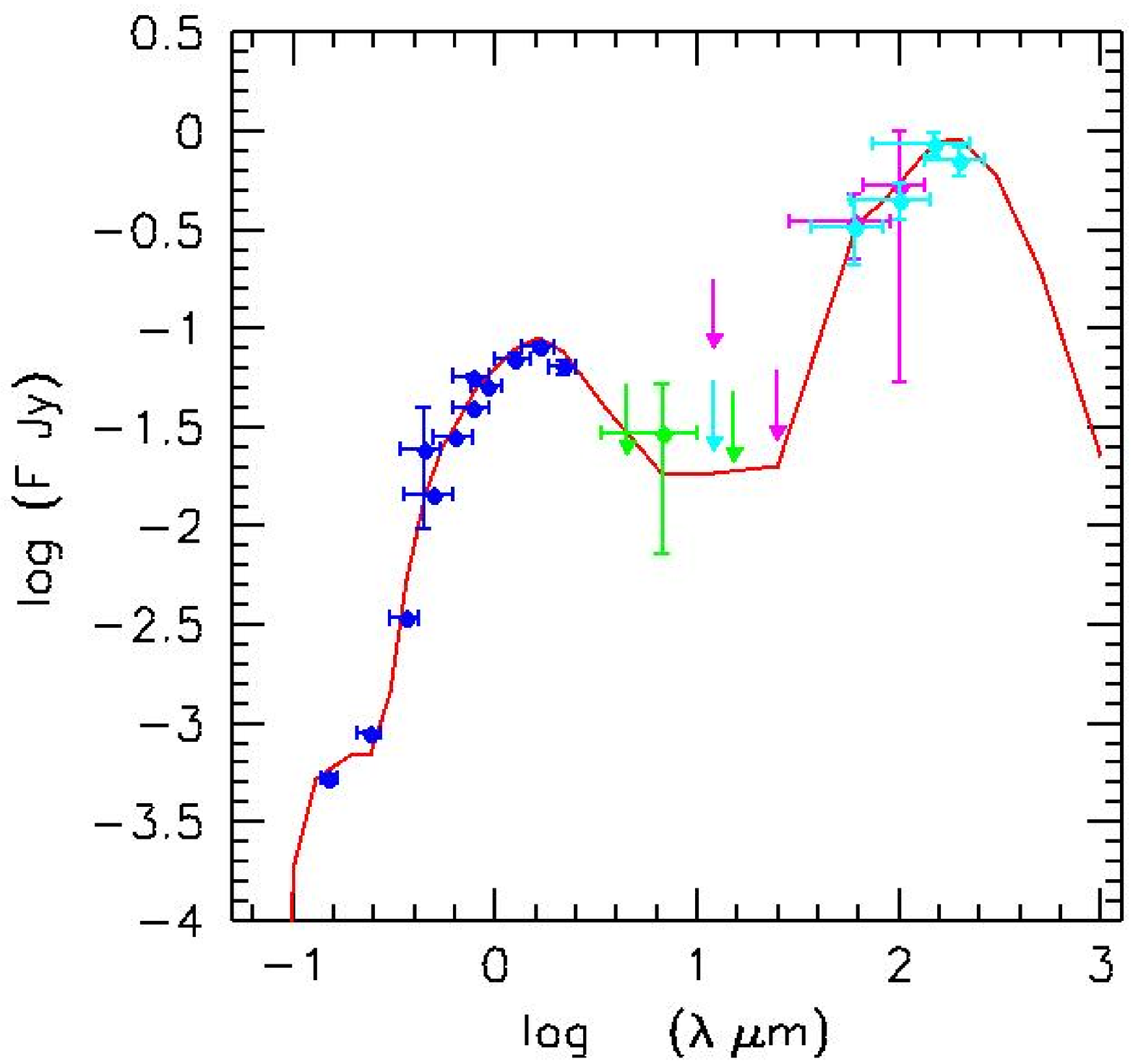}
%
%
\caption{The SEDs of NGC 3593 (left panel)  and NGC 5173 (right panel). Blue circles are total fluxes 
from NED.  Open circles in the
FIR spectral range are from AKARI/FIS Bright Source Catalog \citep{Yama09}
 (black) and IRAS (red)  for NGC 3593.
For NGC 5173, green points are ISO data \citep{Xi}, FIR observations
 are from IRAS (magenta) and  \citet{Temi} (cyan). 
In both panels, the error bars account for band width and 3$\sigma$ uncertainties 
of the fluxes; solid lines are predictions from our chemo-photometric simulations 
at 13.1 and 13.8\,Gyr, respectively.  
The FIR SEDs include two dust components: warm dust (HII regions) and cold dust (heated by the diffuse interstellar field) both 
including PAH molecules.
The warm/cold energy ratio in NGC 3593 is 0.3, similar to the average for 
Spirals, and,  in NGC 5173, double than average for Es,  i.e., 1 instead of 0.5.}
\label{fig:1}       
\end{figure*}

\section{The SPH simulations}
\label{sec:2}

We performed a large set of SPH simulations of galaxy formation starting from collapsing triaxial halos
initially composed of DM and gas. We analyzed  the evolution of  isolated halos in previous works (\cite{CM99, MC}), 
as well as that following from a large grid  of encounters of two halos with different spins, impact parameters, total masses and gas fraction, 
as reported in several previous papers (\cite{MP14a, MP14b, Bet12, Trinetal12}).
All the simulations include self-gravity of
gas, stars and DM, radiative cooling, hydrodynamical pressure,
shock heating, viscosity, SF, feedback both from evolved stars
and type II supernovae, and chemical enrichment.
Simulations provide the synthetic SED, based on evolutionary population synthesis (EPS) models,
at each evolutionary stage. The SED extends
from 0.06 to 1000\,$\mu$m (\cite{M92,M94,M95,M07, S09}, and references therein).
From the grid of physically motivated SPH simulations, we
isolate those simultaneously best fitting the global properties of
selected ETGs, i.e., their absolute $B$-band magnitude, integrated SED,
current morphology and kinematics \citep{MP14a,MP14b}.

The selected simulation  traces a {\it viable} global evolution of the 
galaxy by means of the star formation rate and its related tools, i.e., chemical 
and luminosity evolution, together with the  dynamic evolution of all the galaxy 
components (stars, cold, warm and hot gas, and dark matter).  Our simulations 
allow us to derive the global  properties of the system, in particular  
the spins and relative masses of all the system components, i.e. the encounter/merger 
evolution.

 \begin{figure*}[!]
\centering
\includegraphics[scale=.90]{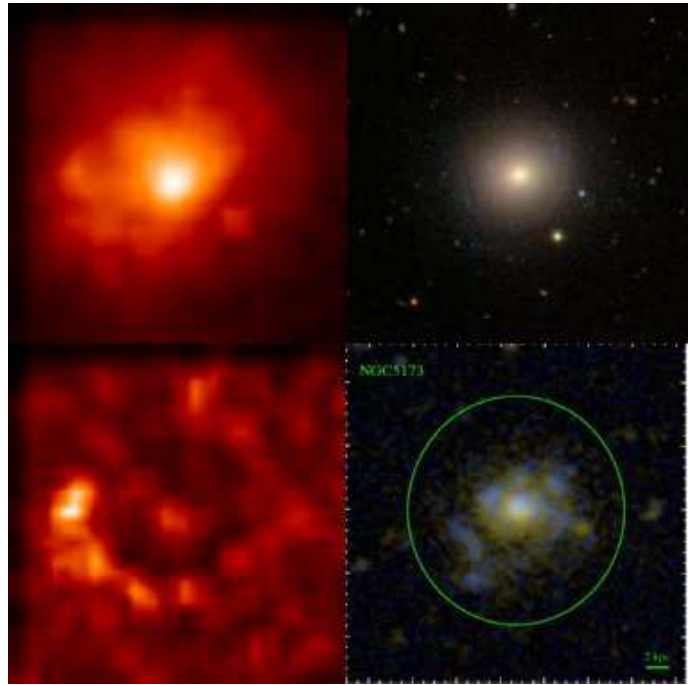}
\caption{  {\it Right:} Top and bottom panels {\bf show} the SDSS $r$-band and the {\it GALEX} 
(FUV blue, NUV yellow) color composite images of NGC 5173, respectively,  {\bf both} on the same 
scale,  2.7'$\times$2.7'.
{\it Left: }Top and bottom  panels  present, 
on the same spatial scale and with resolution 5'' {\bf as the  {\it GALEX}  image,   the maps derived from our simulation}, in the same bands {\bf as in the right panel, at the  age of the galaxy derived from our simulation, i.e., 13.8\,Gyr.}
Both {\bf our} maps have been normalized to the total flux within each simulated image. } 
\label{fig:2}       
\end{figure*}

\section{The counter-rotating galaxy sample with UV emission}
\label{sec:3}

We aim at investigating the characteristics of the largest as possible 
sample of objects with CR components. To this end, we build our sample starting  
from the list of \citet{Gal} and \citet{CB} and mining the current literature  (see e.g. \cite{Sil, K1, K2, KF}). 
 As a final step we checked also the SAURON \citep{Zee02, Em07} and ATLAS-3D \citep{Cap11, Kra11} samples. The only selection criterium is the presence of both stars-gas and/or stars-stars CR. The final sample is composed of 75 galaxies with CR. In order to have an homogeneous data set,  we obtained from the Lyon Meudon Extragalactic Database (LEDA \citet{Pat03}) all the available morphological, photometrical and dynamical data for these galaxies.\\
 Our serendipity selected sample is composed by 80\% of ETGs (morphological type T $\leq$0) and 20\% of late type galaxies. Using the homogeneous distances from LEDA, our sample spans a range of  about five magnitudes in $M_B$, from -22.4 to -17.2 mag, with  $M_B$=-20.04$\pm$1.21 mag in average. Moreover, 
 we exploited the {\it GALEX} archive to obtain far-UV (FUV) and NUV magnitudes. 
We found that   $\sim$50\% of the galaxies in this sample show the presence of a significative FUV emission, a signature of rejuvenation episodes. 

We summarize below the initial conditions, i.e., the relevant characteristics, and  present the results of our SPH simulations for two of our CR-ETGs, NGC 3593 and NGC 5173.

\subsection{NGC 3593} %
This galaxy, classified as SA(s)0/a, is known to host two large counter-rotating 
stellar disks \citep{Bert96} together with a gaseous disk. 
It shows a chaotic pattern of dust patches throughout its main body.
 It is very rich in both neutral (HI) and molecular (H$_2$) gas. 
Their total masses are  5.1$\times$10$^8 M_{\odot}$ and 2.3$\times$10$^8 M_{\odot}$, respectively \citep{Bet01}.
Its $B$-band total absolute magnitude ranges between -16.96 mag to -19.01 mag  accounting for Virgo infall and 3K cosmic microwave background 
corrections, respectively (NED, NASA/IPAC Extragalactic Database http://ned.ipac.caltech.edu),  and adopting H$_0$ = 70\,km\,s$^{-1}$Mpc${^{-1}}$,
$\Omega_{Lambda}$ = 0.73, and $\Omega_{matter}$ = 0.27, as in the HYPERLEDA database  (http://leda.univ-lyon1.fr; \cite{Pat03}).\\
NGC 3593 is a very bright IR source and contains 5.6$\times$10$^5 M_{\odot}$ of dust \citep{Bet01}. 
The galaxy hosts an inner nuclear ring \citep{COM}.  \cite{Coc} suggested that an accretion event occurred between
 2 and 3.6\,Gyr ago, i.e., 1.6\,Gyr after the formation of the main galaxy disk.
We found   that the galaxy has a $FUV-NUV$  color  equal to 1.83,  
and a $NUV-r$ = 4.5,  placing it in the GV.

The global properties of this galaxy, i.e, its SED,  morphology and  $B$-band absolute magnitude, match well with a galaxy encounter  between two equal halos with total mass 8$\times 10^{11}$\,M$_{\odot}$, semi major axis  of 597.2\,kpc and perpendicular spins. The initial conditions of the encounter correspond to a distance of their mass centers of 240\,kpc and to a relative velocity of 33.5\,km\,s$^{-1}$.
 The age of the galaxy we derive, is 13.1\,Gyr and its  $B$-band absolute magnitude, $-18$\,mag.
Figure \ref{fig:1}, left panel,  compares the SED derived from our simulation at the selected snapshot (solid line)
with the total observed fluxes. In this figure JHK fluxes,  corrected for contaminating sources (NED), are reported by \cite{J03},  and IRAC data in the spectral range 3.6 to 8\,$\mu$m,  corrected for saturation effects, are from  \cite{D09}. Our fit is marginally consistent with the data in this spectral range.\\
The stellar mass  predicted inside D$_{25}$ is 1.2$\times$ 10$^{10}$\,M$_{\odot}$. This is in
good agreement with the estimate by \cite{CB}. The average stellar population age, weighted by luminosity,  ranges from 1.3 to 2.4\,Gyr within $r_{eff}$ accounting  for  $B$ or $V$-band luminosities, respectively, in  agreement  with the estimate made by \citet{Coc}.
 The average stellar polupation age within $D_{25}$ is ~7\,Gyr; it becomes younger, 4.4 and 5.3\,Gyr, weighted by $B$ or $V$-band luminosities, 
respectively. The mass of cold gas, i.e., the gas with temperature lower than 20000\,K,  is 3.5$\times 10^{8}$\,M$_{\odot}$ inside the same region. This value is
  within a factor of two from the total gas value, (HI+H2), by \cite{CB} and \cite{Bet01}. \\
  Therefore, the simulation selected  is strongly constrained  by the data since 
  its best-fitting snapshot concurrently well matches  the total $B$-band absolute magnitude,   SED, morphology, cold gas amount of the galaxy and agrees with independent estimates of the average age of its stellar populations and stellar mass.\\
   Figure \ref{fig:CMD} shows the evolutionary path of this  simulation (long-dashed line) in the rest-frame CMD,
 $NUV-r$ versus $r$-band absolute magnitude.  Following its prediction, this galaxy stays on the blue  cloud \citep{Baldry04} 5.7\, Gyr ago.

\subsection{NGC 5173}

This E0 galaxy is likely paired with the SB(rs)b galaxy NGC 5169, separated by 5.5\,arcmin in projection, and by a
velocity difference  of 17 km~s$^{-1}$. The galaxy is a gas-rich ETG since \cite{KF}, which discovered its gas/star CR,  reported a mass of cold gas (HI) of 2.1$\times$10${^9}$\,M$_{\odot}$. 
In the GALEX UV image it is visible a patchy FUV inner ring.\\
The global properties of this galaxy match well with a galaxy encounter 
between two equal halos with total mass 4$\times 10^{12}$\,M$_{\odot}$, semi major axis 1\,Mpc and parallel spins.
The initial conditions of the encounter correspond to a distance of their mass centers of 889\,kpc and to a relative velocity of 62\,km\,s$^{-1}$.\\
From this simulation we derive that the age of the galaxy is 13.8\,Gyr, and  its $B$-band absolute magnitude -20.1 mag, to be compared with  the value of -19.72$\pm$0.41  mag from HyperLeda catalog.
The predicted SED is compared with observations in Fig. \ref{fig:1}.
 The galaxy has a  blue {\it GALEX} UV color, FUV-NUV = 0.57,  and a blue UV-optical color,
 NUV-r = 3.73, placing also this ETG in the GV.\\
Fig. \ref{fig:2}  compares both the UV composite {\it GALEX} image and the $r$-band SDSS one with our simulated maps.  These are on the same bands and
 spatial scale as the observed ones.  The spatial resolution of the simulated maps is 5'',  the same as the {\it GALEX} image (right bottom panel).
The mass of cold gas,  i.e.,  that of the gas with temperature lower than 20000\,K,  is 1.2$\times 10^{9}$\,M$_{\odot}$ inside its  $D_{25}$,
in good agreement with previous estimates \citep{KF}.

\begin{figure*}[!]
 \begin{center}
\includegraphics[scale=.75]{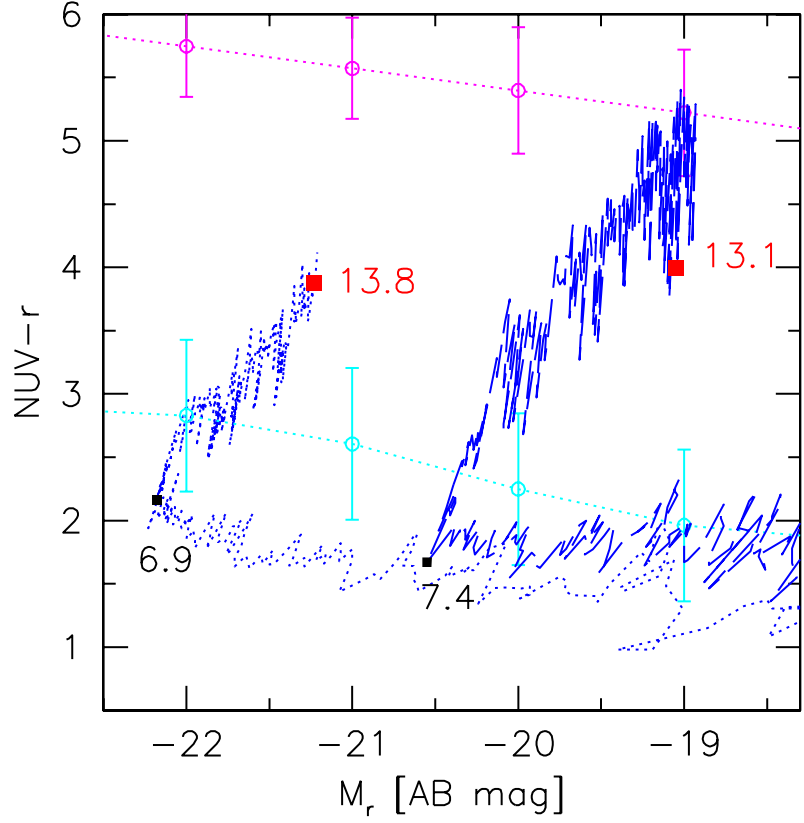}
\end{center}
\caption{ Rest-frame $NUV-r$ versus $M_r$ CDM of the simulations which best fit of the global properties of NGC~3593 (long-dashed line) and NGC~5173 (dotted line). 
The red filled squares correspond to the current colors of these ETGs, both  in the GV, well matched by the simulations stopped at  14\,Gyr. 
The evolutionary path of both galaxies shows the transition from the blue toward the red sequence through the GV. Numbers are the ages in Gyr 
 along the evolutionary traces, starting from bottom right. The red sequence (magenta) and the blue cloud (cyan) are plotted following prescriptions in \citet{Wyderetal07}.}
\label{fig:CMD}       
\end{figure*}

Figure \ref{fig:CMD} shows the evolutionary path of this  simulation (dotted line) in the rest-frame CMD,
 $NUV-r$ versus $r$-band absolute magnitude. This galaxy lived in the blue cloud 6.9\,Gyr, and, in the same time,  through the GV to reach its actual position.
%

\section{Conclusions}
\label{sec:4}
We have assembled a sample of 75 CR-ETGs and studied their UV properties from 
{\it GALEX} archive. We found that about 50\% of them show strong FUV emission. This is a percentage
significantly larger than that found for a general sample of ETGs \citep[30\%, see e.g.][]{SK07}. 
On this basis we are performing  a detailed analysis of the properties of the whole sample making use of SPH simulations  with chemo-photometric implementation which give us insight into their evolution. 
The  galaxies studied  here,  NGC 3593  and NGC 5173, are both   ETGs. They show rather blue colors in the  rest-frame $NUV-r$ versus $r$-band absolute magnitude CMD,  placing them in the GV instead of on the red sequence as expected for the majority of ETGs \citep[e.g.][]{MP14a, MP14b}. 
Both galaxies  are gas rich and NGC 5173 shows a   patchy structure in the UV {\it GALEX} bands.
Our chemo-photometric SPH simulations allow us to trace the evolutionary path of these galaxies in a fully consistent way. 
Their global properties, i.e, their $B$-band total absolute magnitudes,  SEDs,  morphologies and mass of cold gas, match well with simulations that correspond to
 a galaxy encounter between two equal mass halos. Their initial total masses are 8$\times$10$^{11}$\,M$_{\odot}$ and 5 times more for NGC 3593 and NGC 1573, respectively.
 Their current ages are   13.1 and 13.8\,Gyr, respectively. 
Their evolution in the rest-frame $NUV-r$ versus $r$-band absolute magnitude  CMD is marked by several rejuvenation episodes, which place them,  at their current age,
 outside the red sequence, where normal ETGs stay.  Their position in the GV emphasises
a  quite different evolutionary path of these galaxies, 
as  traced by our simulations. These simulations point out their origin in a galaxy encounter rather than a merger, as in the majority of ETGs analyzed with the same approach \citep{MP14a, MP14b}.    
We plan to match the  global properties  of the whole sample of CR galaxies here presented. Our simulations with chemo-photometric implementation,  will allow us to give further insight into the evolution of ETGs, in particular of those with CR.
 We  will aim at answering to the question concerning the mechanism of their formation and the origin of CR by comparing the properties of CR galaxies in our sample with those of galaxies without CR.

%

%
\acknowledgments
We acknowledge the partial financial support by contract ASI-INAF  
I/009/10/0, and the partial financial support by contract INAF/PRIN 2011 ``Galaxy 
Evolution with the VLT Survey Telescope (VST)" (P.I.; A. Grado).

%

\end{document}